\documentclass[11pt, a4paper]{article}   
\usepackage[T1]{fontenc}
\usepackage[utf8]{inputenc}
\usepackage{authblk}
\usepackage{color}
\usepackage{graphicx}
\usepackage{geometry} 
\usepackage{multirow}
\usepackage{marvosym} 
\definecolor{purple}{rgb}{0.5,0,0.5}

\title{\textbf{Design of the Front End Electronics for the Infrared Camera of JEM-EUSO, and manufacturing and verification of the prototype model}}   
\author[a]{{\Large Óscar Maroto}}
\author[a]{{\Large Laura Díez-Merino}}
\author[a]{{\Large Jordi Carbonell}}
\author[a]{{\Large Albert Tomàs}}
\author[b]{{\Large Marcos Reyes}}
\author[b]{{\Large Enrique Joven}}
\author[b]{{\Large Yolanda Martín}}
\author[c]{{\Large J. A. Morales de los Ríos}} 
\author[c]{{\Large \\Luis Del Peral}} 
\author[c,b]{{\Large M. D. Rodríguez Frías}}
\affil[a]{{\footnotesize \textit{SENER, C/ Creu Casas i Sicart, 86-88, Parc de l’Alba 08290 Cerdanyola del Vallès, Barcelona, Spain}}}
\affil[b]{{\footnotesize \textit{Instituto de Astrofísica de Canarias (IAC), Vía Lactea s/n, Tenerife, Spain}}}
\affil[c]{{\footnotesize \textit{SPace and AStroparticle Group (SPAS), UAH, Madrid, Spain}}}

\date{\today}    
\begin{document}
\maketitle
\begin{abstract}
The Japanese Experiment Module (JEM) Extreme Universe Space Observatory (EUSO) will be launched and attached to the Japanese module of the International Space Station (ISS). Its aim is to observe UV photon tracks produced by ultra-high energy cosmic rays developing in the atmosphere and producing extensive air showers.

The key element of the instrument is a very wide-field, very fast, large-lense telescope that can detect extreme energy particles with energy above $10^{19}$ eV. The Atmospheric Monitoring System (AMS), comprising, among others, the Infrared Camera (IRCAM), which is the Spanish contribution, plays a fundamental role in the understanding of the atmospheric conditions in the Field of View (FoV) of the telescope. It is used to detect the temperature of clouds and to obtain the cloud coverage and cloud top altitude during the observation period of the JEM-EUSO main instrument.

SENER is responsible for the preliminary design of the Front End Electronics (FEE) of the Infrared Camera, based on an uncooled microbolometer, and the manufacturing and verification of the prototype model.

This paper describes the flight design drivers and key factors to achieve the target features, namely, detector biasing with electrical noise better than $100 \mu$V from $1$ Hz to $10$ MHz, temperature control of the microbolometer, from $10^{\circ}$C to $40^{\circ}$C with stability better than $10$ mK over $4.8$ hours, low noise high bandwidth amplifier adaptation of the microbolometer output to differential input before analog to digital conversion, housekeeping generation, microbolometer control, and image accumulation for noise reduction.

It also shows the modifications implemented in the FEE prototype design to perform a trade-off of different technologies, such as the convenience of using linear or switched regulation for the temperature control, the possibility to check the camera performances when both microbolometer and analog electronics are moved further away from the power and digital electronics, and the addition of switching regulators to demonstrate the design is immune to the electrical noise the switching converters introduce.

Finally, the results obtained during the verification phase are presented: FEE limitations, verification results, including FEE noise for each channel and its equivalent NETD and microbolometer temperature stability achieved, technologies trade-off, lessons learnt, and design improvement to implement in future project phases.
\end{abstract}


\section{Introduction}             
JEM-EUSO (Extreme Universe Space Observatory on the Japanese Experiment Module)\cite{Takahashi2009}, \cite{Ebisuzaki2014}, \cite{Adams2013} is a new type of observatory that will utilize very large volumes of the Earth’s atmosphere as a detector of the most energetic particles in the Universe. Its aim is to observe UV photon tracks produced by Ultra High Energy Cosmic Rays (UHECR), with energy above $10^{19}$ eV, generating Extensive Air Showers (EAS) in the atmosphere. The Atmospheric Monitoring System (AMS)\cite{Neronov2011} plays a fundamental role in our understanding of the atmospheric conditions in the Field of View (FoV) of the telescope and includes an infra-red camera for cloud coverage and cloud top height detection.

The monitoring of the cloud coverage by JEM-EUSO with an Atmospheric Monitor System is crucial to estimate the effective exposure with high accuracy and to increase the confidence level in the UHECRs events in particular at the threshold energy of the telescope. The AMS is a system used to detect the temperature of clouds and to obtain the cloud coverage and cloud top altitude during the observation period of the JEM-EUSO main instrument. Cloud top height retrieval can be performed using either stereo vision algorithms (therefore, two different views of the same scene are needed) or accurate radiometric information, since the measured radiance is basically related to the target temperature and therefore, according to standard atmospheric models, to its altitude. The observed radiation is basically related to the target temperature and emissivity and, in this particular case, it can be used to get an estimate of the cloud heights. The AMS will comprise an Infrared Camera (IRCAM)\cite{Rodriguez-Frias2013},\cite{Morales2013},\cite{JEM-EUSO2014},\cite{Morales2014}, a LIDAR and JEM-EUSO slow data.

IRCAM, the technological contribution of the Spanish Consortium, is able to detect infrared radiation of the target with emissivity ($\epsilon$) greater than 0.6 and lower than 1 and estimates the temperature of the target under investigation with accuracy better than $3$ K. IRCAM consists of three subsystems:
\begin{itemize}
\item IRCAM Telecope Assembly.
\item IRCAM Electronics Assembly.
\item IRCAM Calibration Unit.
\end{itemize}

The main function of the IRCAM Telescope assembly is to acquire the infrared radiation and to convert it into digital counts. For that purpose, the Telescope Assembly includes the IRCAM detector, an uncooled microbolometer, the dedicated electronics to control it, the Front End Electronics (FEE), and the optics.

The IRCAM Electronics Assembly provides mechanisms to process and transmit the images obtained by the FEE from the microbolometer. It is composed of the Instrument Control Unit (ICU) and the Power Supply Unit (PSU). Their main function is to control and manage the overall system behavior, including the data management, the power drivers and the mechanisms. 

Due to the requested accuracy measurement the IRCAM will perform an On-Board calibration by means of a dedicated Calibration Unit.

SENER is responsible for the preliminary design of the Front End Electronics (FEE) within the IRCAM, and the manufacturing and verification of the prototype model, under IAC supervision.

\section{IRCAM FEE Main Requirements}             

The main requirements the IRCAM FEE shall comply with are listed in the following:
\begin{enumerate}
\item	IRCAM FEE shall acquire images from an infrared detector, type ULIS UL 04 17 1, covering the spectral range $8-14 \mu$m and measuring temperatures between $200$ K and $320$ K\cite{ULIS2014}.
\item	IRCAM FEE shall generate the bias needed by the infrared detector to operate. These are enlisted in Table \ref{Table:requirements}.
\item	The detector shall be controlled in temperature between $10^{\circ}$ C and $40^{\circ}$ C with stability better than $10$ mK over $4.8$ hours, by commanding the Thermo-Electric Cooler (TEC) of the microbolometer.
\item	The NETD of the IRCAM FEE shall be lower than $75$ mK $\MVAt$ $300$ K and optics F\#1 of the target.
\item	IRCAM FEE shall be able to perform frame averaging of at least 4 consecutive frames.
\item 	FEE shall be able to perform operation of sum and subtracting of frames.
\item 	IRCAM FEE shall be composed of two different modules, each of them, located at different positions, joint by a flex cable:
\begin{itemize}
\item	FPA Focal Plane assembly (detector + mechanical I/F to the Optical subsystem).
\item	FEE electronic box.
\end{itemize}
\end{enumerate}
\section{IRCAM FEE Flight Design Overview}             
IRCAM FEE has been divided into two different modules, namely, the Focal Plane Array (FPA) containing the low noise electronics needed for the infrared detector to operate optimally, and the Front End Electronics (FEE) hosting the digital electronics and TEC control circuitry.

\begin{figure}[!tb] 
\begin{center}
\includegraphics[width=14cm]{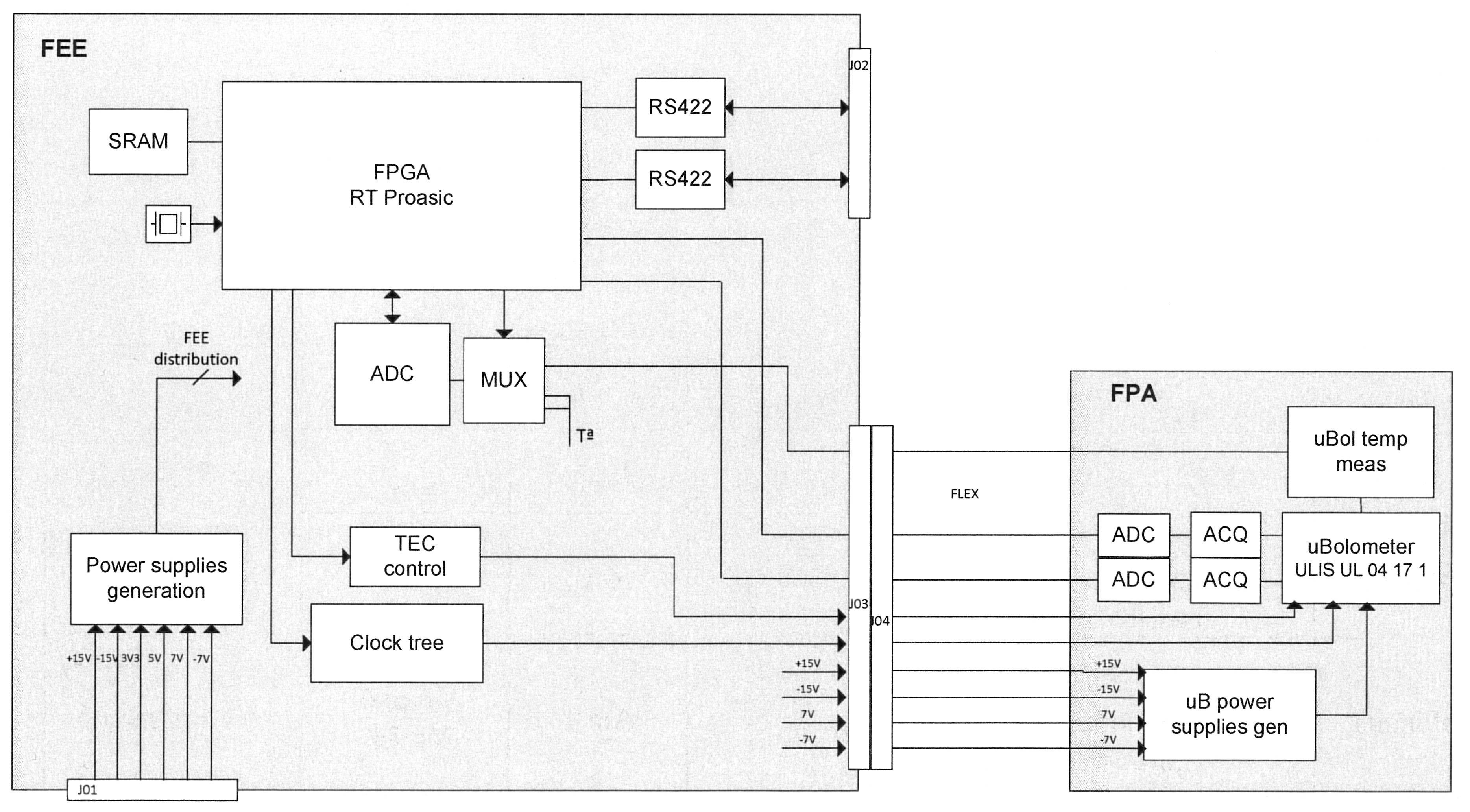}
\caption[IRCAM FEE block diagram]{IRCAM FEE block diagram}
\label{Fig:FEE}
\end{center}
\end{figure}

\begin{table}[!h]
\caption[Requirements for microbolometer power supplies.]{Requirements for microbolometer power supplies.} \label{Table:requirements}
\begin{center}
\begin{tabular}{|p{28mm}|p{14mm}|p{18mm}|p{20mm}|p{20mm}|p{40mm}|}
\hline
\textbf{Electrical }   & \textbf{Bias} & \textbf{Optimum} & \textbf{Range} & \textbf{Maximum} & \textbf{Maximum} \\
\textbf{function name}   & \textbf{Type} & \textbf{value} & \textbf{value} & \textbf{current} & \textbf{RMS noise} \\
   &  & \textbf{@ 300K} &  &  &  \\
\hline
\hline
VDDA             &	Fixed &	$5$ V        & 	 & 	$60$ mA	 & $< 100 \mu$V \\
(analog supply)  &	      & $\pm 100$ mV & 	 &  & \\
\hline
VDDL &	Fixed &	$3.3$ V  &	& 	$5$ mA	& $< 100$ mV \\
(digital supply) &	 &	$\pm 300$ mV &	& 	&  \\
\hline
VBUS&	Fixed &	$2.8$ V  &  &	 $1$ mA & 	$< 100 \mu$V \\
(microbolometer biasing) &	 &	$\pm 25$ mV &  &	  & 	 \\
\hline
GFID &	Tunable & Given &	$0$ to $5$V  &	$1$ mA & 	$2 \mu$V (1Hz to 1kHz) \\
(microbolometer &	 & in STR &	 $\pm 5$ mV &	 & 	$5 \mu$V (1Hz to 10kHz) \\
biasing) &	 &  &	  &	 & 	$100 \mu$V (1Hz to 10MHz) \\
\hline
VSK (blind &	Tunable &	Given &	$2.0$ to $5.5$ V  &	$1$ mA &	 $2 \mu$V (1 Hz to 1 kHz) \\
microbolometer & &	in STR &	 $\pm 5$mV &	 &	 $5 \mu$V (1 Hz to 10 kHz) \\
biasing) &	 &	 &	 &	 &	 $100 \mu$V (1 Hz to 10 MHz) \\
\hline
GSK (blind &	Fixed &	$2.2$ V  &	& $1$ mA &	$2 \mu$V (1 Hz to 1 kHz) \\
microbolometer &	 &	$\pm 50$ mV & &	 &	$5 \mu$V (1 Hz to 10 kHz) \\
biasing) &	& &	 &	 &	$100 \mu$V (1 Hz to 10 MHz) \\
\hline
\end{tabular}
\end{center}
\end{table}

Figure 1 shows the block diagram of the IRCAM FEE electronics.The main functionalities implemented in the IRCAM FEE are:

\subsection{Power supplies generation for the FEE PCB}             

All the power supplies are cold redundant and consequently OR-ing is needed. This block also filters the input power supplies and generates secondary voltages by using linear regulators and point of loads.

\subsection{TEC temperature control}             
\begin{figure}[!tb] 
\begin{center}
\includegraphics[width=14cm]{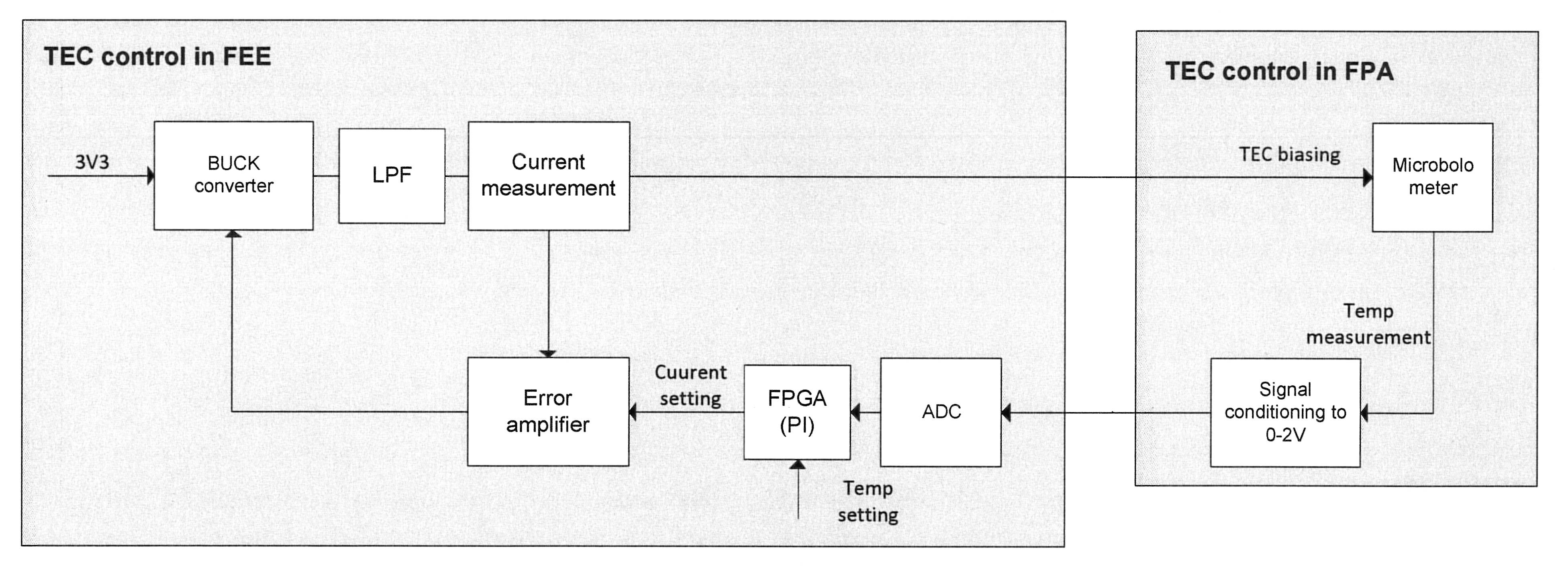}
\caption[TEC control block diagram.]{TEC control block diagram. \label{Fig:TEC}}
\end{center}
\end{figure}

The microbolometer performance is optimum when the detector is stabilized to a constant temperature $\pm 10$ mK for the image acquisition time, presently defined as $4.8$ h.
To fulfil this requirement, the design shown in Figure \ref{Fig:TEC} is featured in the IRCAM FEE.

This circuitry consists of a current regulator implemented with a buck converter and controlled by a double loop. The possible setting temperatures are between $10$ and $40^{\circ}$C, and  it is only possible to heat up due to the facts that the temperature of the FPA is expected to be lower than the microbolometer setting temperature, and in case of heating/cooling cycles were mixed during image acquisition, the thermal stability would be much worse than $10$ mK.

The advantage of using a buck converter to implement the current regulator is the efficiency of the circuit, and as the power availability is little, it has been decided to choose this option. The drawback is that the switching noise could affect the low noise circuits as microbolometer power supplies generation or image acquisition. The alternative would be to use a linear regulator that is better in terms of noise, but the power consumption for big temperature jumps would be unaffordable. To analyze it, in the prototype the two options have been implemented to have a trade-off of both technologies.

\subsection{Optical detector power supplies biasing}             
The required power supplies for the optical detector have very severe noise specifications and due to this, most of the power supplies generators are located near the optical sensor.
\begin{figure}[!tb] 
\begin{center}
\includegraphics[width=14cm]{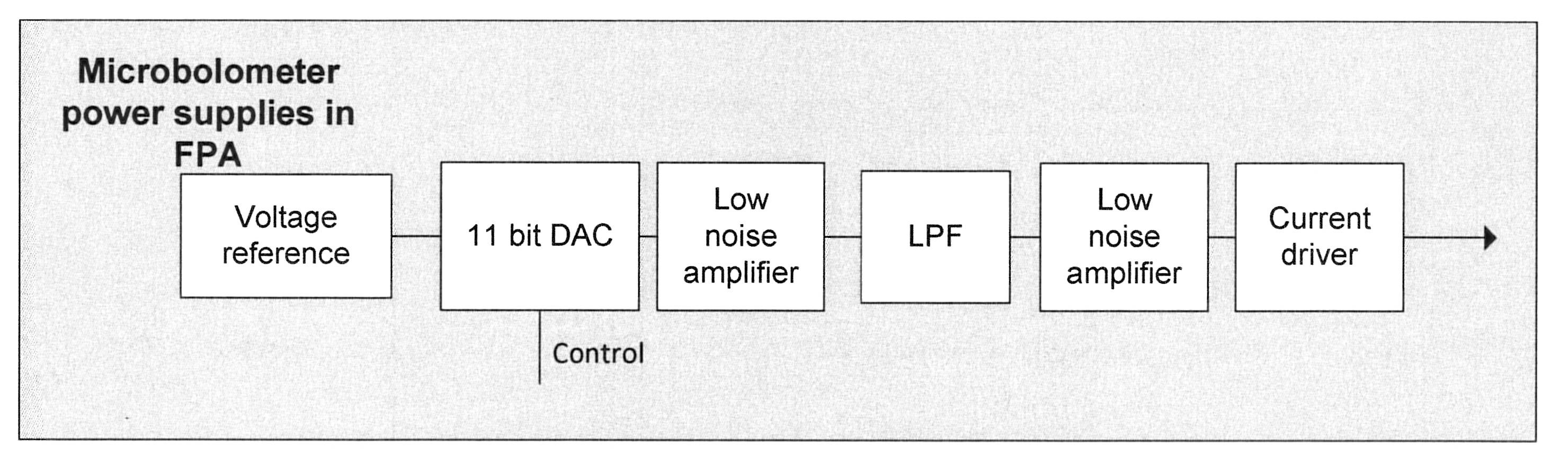}
\caption[Microbolometer power supplies bias.]{Microbolometer power supplies bias. \label{Fig:microbolometer}}
\end{center}
\end{figure}

The architecture of the optical detector power supplies generation is based on a voltage reference that is adjusted and filtered by using very low noise operational amplifiers in non-inverting or inverting configuration. To filter the noise, a Sallen-key topology has been selected. The current driver only applies to VDDA, which delivers $60$ mA maximum, and consists of a bipolar transistor working in linear zone, controlled by an operational amplifier output. 

GFID and VSK power supplies must be tunable voltages. To provide the design with this capability, an 11bit DAC has been selected. $11$ bits means a resolution of $2.7$ mV in VSK and $2.5$ mV in GFID, thus complying with the specification. The DAC used has a simple R-2R topology that minimizes the noise.

\subsection{Optical detector data acquisition}             
This block is in charge of amplifying the signal provided by the microbolometer and performing the analog to digital conversion, to maximize signal to noise ratio. The microbolometer provides data to each channel at $10$ Msamples/s and therefore, the analog acquisition chain has been designed to cope with this data throughput.
\begin{figure}[!tb] 
\begin{center}
\includegraphics[width=14cm]{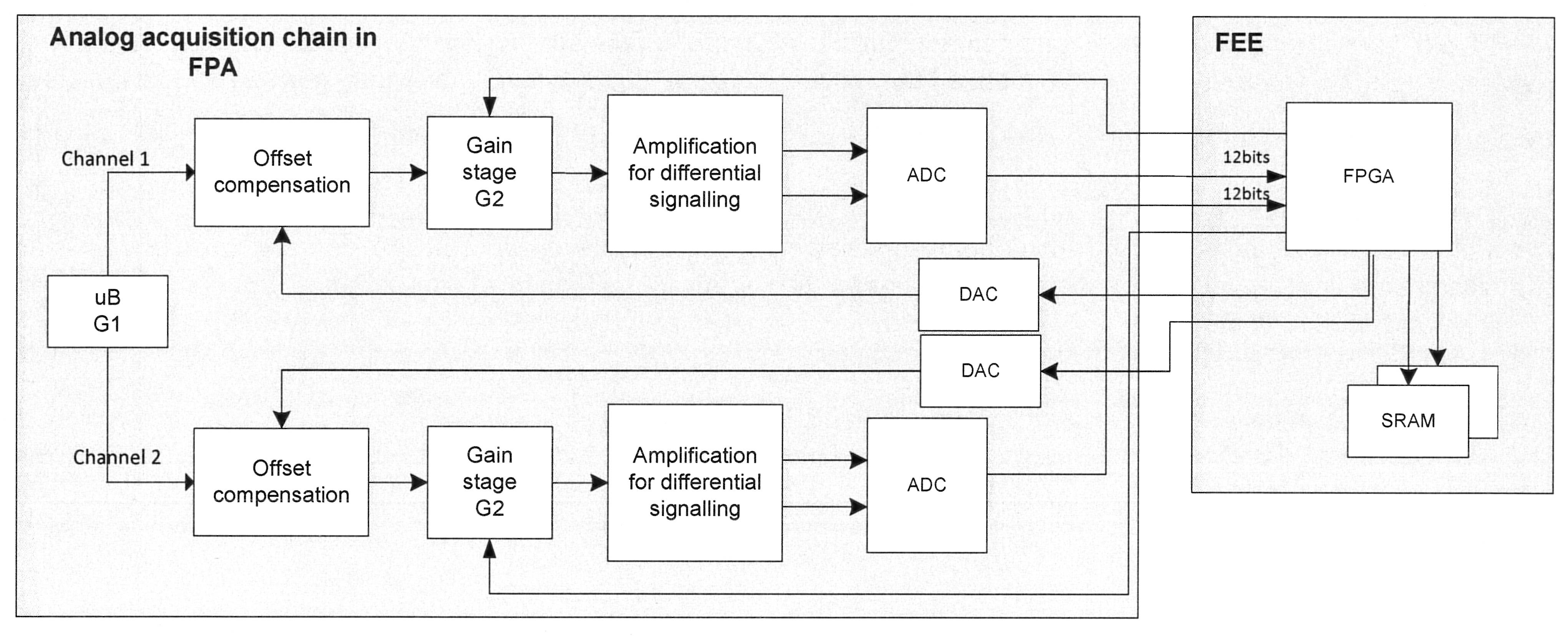}
\caption[Analog acquisition chain block diagram.]{Analog acquisition chain block diagram. \label{Fig:acquisition}}
\end{center}
\end{figure}

The microbolometer measures the cloud temperature in a range between $200$ K and $320$ K. The typical responsivity of the microbolometer is specified by the manufacturer as $5$ mV/K, but it can be modified to obtain different images when the f-number of the optics changes. The range within it can be modified has been considered between $1.5$ mV/K and $8$ mV/K. The acquisition chain has been designed to allow a programmable offset and gain to adjust the input signal to the voltage required by the ADC, independently of the responsivity. 

The offset shall be programmable through the FPGA. It generates a digital voltage, which is converted using a DAC and subtracted to the input signal in the first stage of the analog chain. The value of the offset is: 
\begin{equation}
V_{offset} = T_{min} K_{\mu B} G1
\end{equation}
where $T_{min}$ is the minimum detectable temperature ($200$ K);
$K_{\mu B}$, the responsivity of the microbolometer; and $G1$ is the gain of the microbolometer.

The whole gain of the analog chain is the product of the internal microbolometer gain ($G1$) and the gain of the second stage of the chain ($G2$).  The gain $G1$ is programmable via serial bus in the $\mu B$ and the gain $G2$ is programmable via a resistance ladder multiplexed in the feedback loop. The product $G1\cdot G2$ must amplify the voltage input signal to a $1 V_{pp}$ differential signal (expected voltage at the ADC input), it is:
\begin{equation}
G1\cdot G2 = \frac{V_{in-ADC}}{V_{max-\mu B}} = \frac{V_{in-ADC}}{(T_{max}-T_{min})K_{\mu B}}
\end{equation}

The possible values for $G1$ are specified by the manufacturer and the values for $G2$ have been calculated to minimize the error in the amplification. These are detailed in Table \ref{Table:gain}. All combinations between $G1$ and $G2$ are allowed to adapt the chain gain to the adjustable responsivity of the microbolometer. Minimum gain allowed shall be $1$, while maximum gain for very low power signals could rise up to $6.705$.
\begin{table}[!h]
\caption[Programmable gain values in the analog acquisition chain]{Programmable gain values in the analog acquisition chain \label{Table:gain}}
\begin{center}
\begin{tabular}{|p{1.2cm}|p{1.2cm}|}
\hline
\textbf{G1}   & \textbf{G2}   \\
\hline
\hline
1.000 &	1.000 \\
1.125 &	1.070  \\
1.290 &	1.140 \\
1.500 &	1.210 \\
1.800 &	1.280 \\
2.250 &	1.350 \\
3.000 &	1.420 \\
4.500 &	1.490 \\
\hline
\end{tabular}
\end{center}
\end{table}

Finally, after offset compensation and gain stage, the signal shall be converted from single-ended to differential signal before entering in the ADC.

The noise analysis performed for this module, which is the main contributor to the NETD added by the FEE is depicted in Table \ref{Table:analog}. 
\begin{table}[!h]
\caption[Analog acquisition chain noise for different responsivity values (microbolometer noise not included).]{Analog acquisition chain noise for different responsivity values (microbolometer noise not included).} 
\label{Table:analog}
\begin{center}
\begin{tabular}{|p{30mm}|p{30mm}|p{25mm}|p{40mm}|}
\hline
\textbf{Responsivity }   & \textbf{Analog Chain} & \textbf{Total Noise} & \textbf{Equivalent NETD} \\
\textbf{[mV/K]}   & \textbf{parameters} & \textbf{[$\mu$V]} & \textbf{[mK]} \\
\hline
\hline
K=1.50           &	G1=4.50        & 183.99	 & 	122.66 \\
                 &	G2=1.21        & 	 & 	 \\
\hline
K=2.54           &	G1=3.00        & 179.95	 & 	70.85 \\
                 &	G2=1.07        & 	 & 	 \\
\hline
K=3.71           &	G1=1.50        & 192.75	 & 	51.95 \\
                 &	G2=1.49        & 	 & 	 \\
\hline
K=4.36           &	G1=1.29        & 190.39	 & 	43.67 \\
                 &	G2=1.42        & 	 & 	 \\
\hline
K=5.22           &	G1=1.29        & 183.93	 & 	35.24 \\
                 &	G2=1.21        & 	 & 	 \\
\hline
K=6.40           &	G1=1.29        & 178.09	 & 	27.82 \\
                 &	G2=1.00        & 	 & 	 \\
\hline
K=7.30           &	G1=1.00        & 181.88	 & 	24.91 \\
                 &	G2=1.14        & 	 & 	 \\
\hline
\end{tabular}
\end{center}
\end{table}

Considering the microbolometer has a NETD around $60$ mK for high responsibility values, the system NETD would be increased from $60$ mK to $64.97$ mK, only $5$ mK, by using equation:
\begin{equation}
NETD_{IRCAM-FEE} = \sqrt{NETD_{FEE}^2+NETD_{\mu B}^2}
\end{equation}

\subsection{Digital design}             
The control of the whole system has been implemented by means of a RT Proasic FPGA, being it radiation tolerant up to $30$ krads that ensures the system is under known conditions regardless of the radiation events received. This electronics will be the responsible of the following activities:
\begin{itemize}
\item	Acquiring images from the microbolometer.
\item	Processing the images and storing them into memory.
\item	Generating the clock signals for the microbolometer.
\item	Generating the power set-ups required by the microbolometer.
\item	Generating the signalling for the FPA and FEE electronics.
\item	Transmitting the images to the ICU.
\item	Receiving TC from the ICU.
\item	Sending TM to the ICU.
\item	Managing the IRCAM FEE modes.
\item	Acquiring instrument temperature sensors. 
\end{itemize}

\begin{figure}[!tb] 
\begin{center}
\includegraphics[width=14cm]{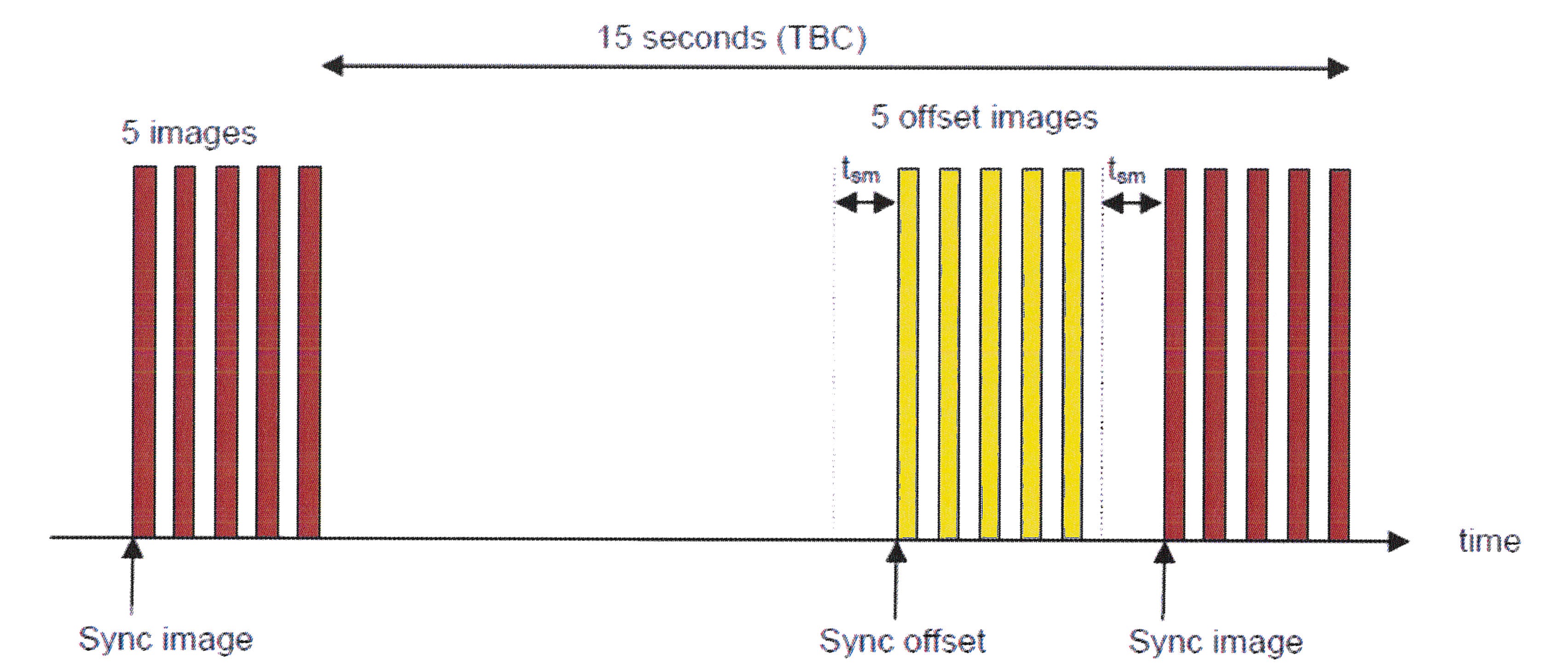}
\caption[Image acquisition strategy]{Image acquisition strategy. \label{Fig:figure5}}
\end{center}
\end{figure}

One of the key points of the digital design is the acquisition strategy to receive, store, process and retrieve an image. The image acquiring strategy is shown in Figure \ref{Fig:figure5}. 

Five data and offset images will be acquired and reduced within $15$ seconds. The image reduction consists on discarding the first image acquired and averaging the other four images. Offset reduction is similar, first image will be discarded and the other four images will be averaged. Prior to the final image retrieval, the offset image is subtracted from the data image.

\subsection{EEE parts selection}             
IRCAM FEE preliminary design has been carried out taking into account flight standards, being the components radiation tolerant to $30$ krads total dose radiation, and immune to LET under $60$ MeV cm$^2$/mg . Apart from the radiation tolerance feature, the electronics components have been selected according to the following rules:
\begin{itemize}
\item	ESCC class B components for those under ESA standards.
\item	QML V components for microcircuits under MIL standards.
\item	JANS for diodes and transistors under MIL standards.
\item	EFR-R for passives under MIL standards.
\end{itemize}

\subsection{Mechanical design}             
\begin{figure}[!tb] 
\begin{center}
\includegraphics[width=14cm]{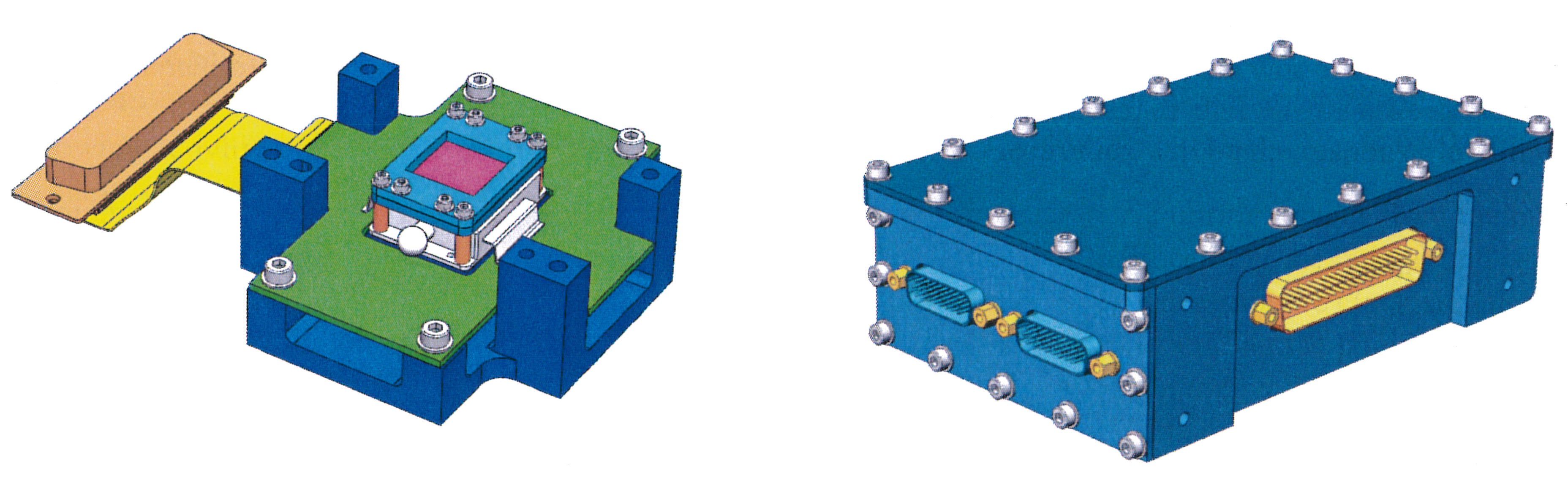}
\caption[FPA and FEE mechanical design.]{FPA and FEE mechanical design. \label{Fig:figure6}}
\end{center}
\end{figure}

The mechanical design for the FPA and FEE can be seen in Figure \ref{Fig:figure6}

The FPA consists of the following elements:
\begin{itemize}
\item	Frame and supports to host the optical filter.
\item	Heat conductor allowing heat conduction of the optical detector and associated electronics to a cold plate.
\item	Rigid-flex PCB, with the needed low noise electronics, such as microbolometer power supplies generation, image data amplification and A/D conversion.
\end{itemize}

On the other hand, FEE components are:
\begin{itemize}
\item	FEE mechanical box.
\item	FEE PCB.
\end{itemize}

\begin{figure}[!tb] 
\begin{center}
\includegraphics[width=14cm]{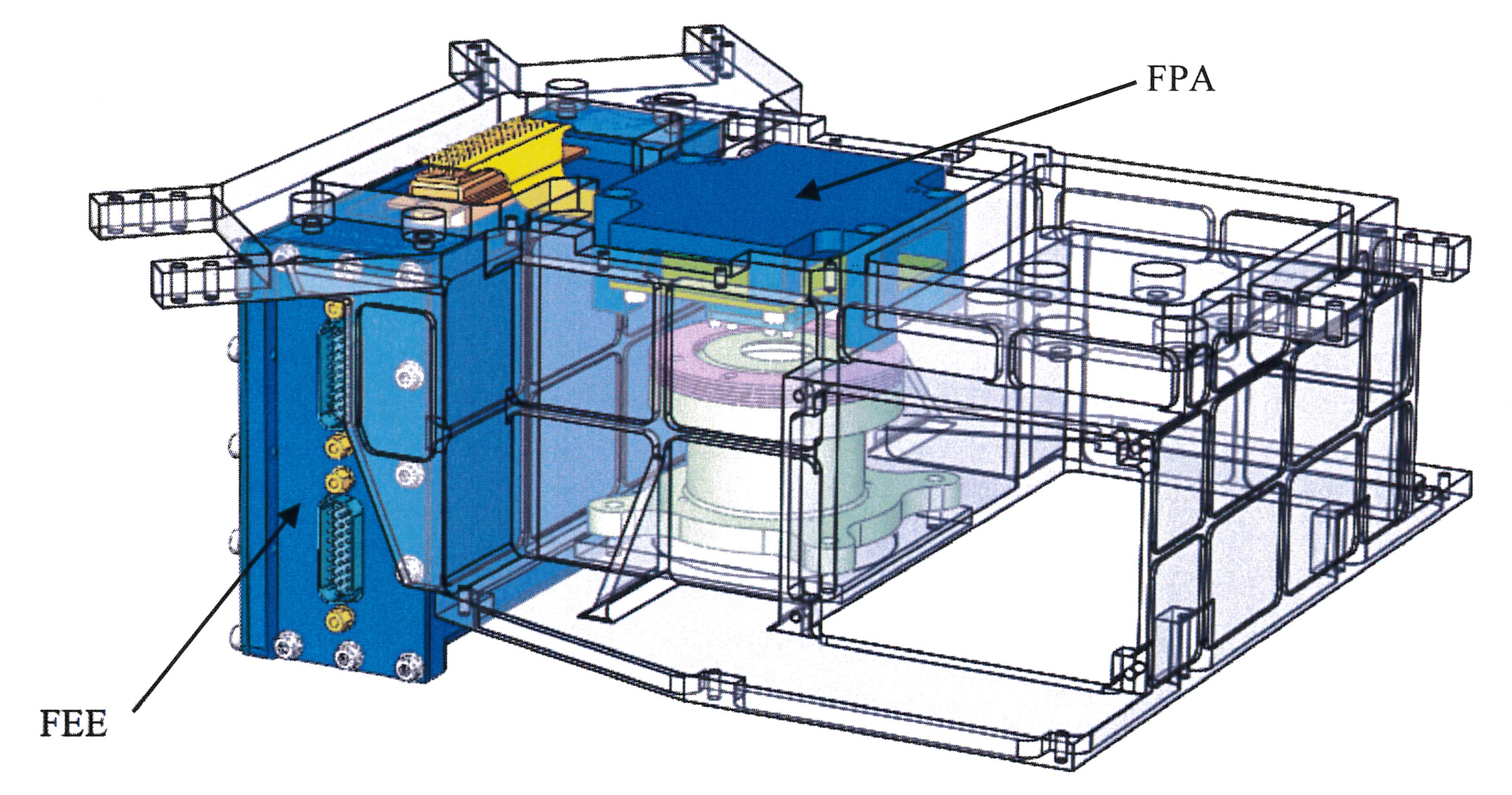}
\caption[IRCAM FEE location within Telescope Assembly.]{IRCAM FEE location within Telescope Assembly. \label{Fig:figure7}}
\end{center}
\end{figure}

The IRCAM FEE is hosted within the Telescope Assembly as depicted in Figure \ref{Fig:figure7}.

\section{IRCAM FEE Prototype Design and Manufacturing}             
The aim of the IRCAM FEE prototype design (FEEP) is to have a functional IRCAM FEE model complying with the requirements described in section 2 with commercial components equivalent to those used for the flight design. A microbolomoter type ULIS UL 04 17 1 delivered by IAC has also been mounted to allow full FEEP verification. 

\begin{figure}[!tb] 
\begin{center}
\includegraphics[width=14cm]{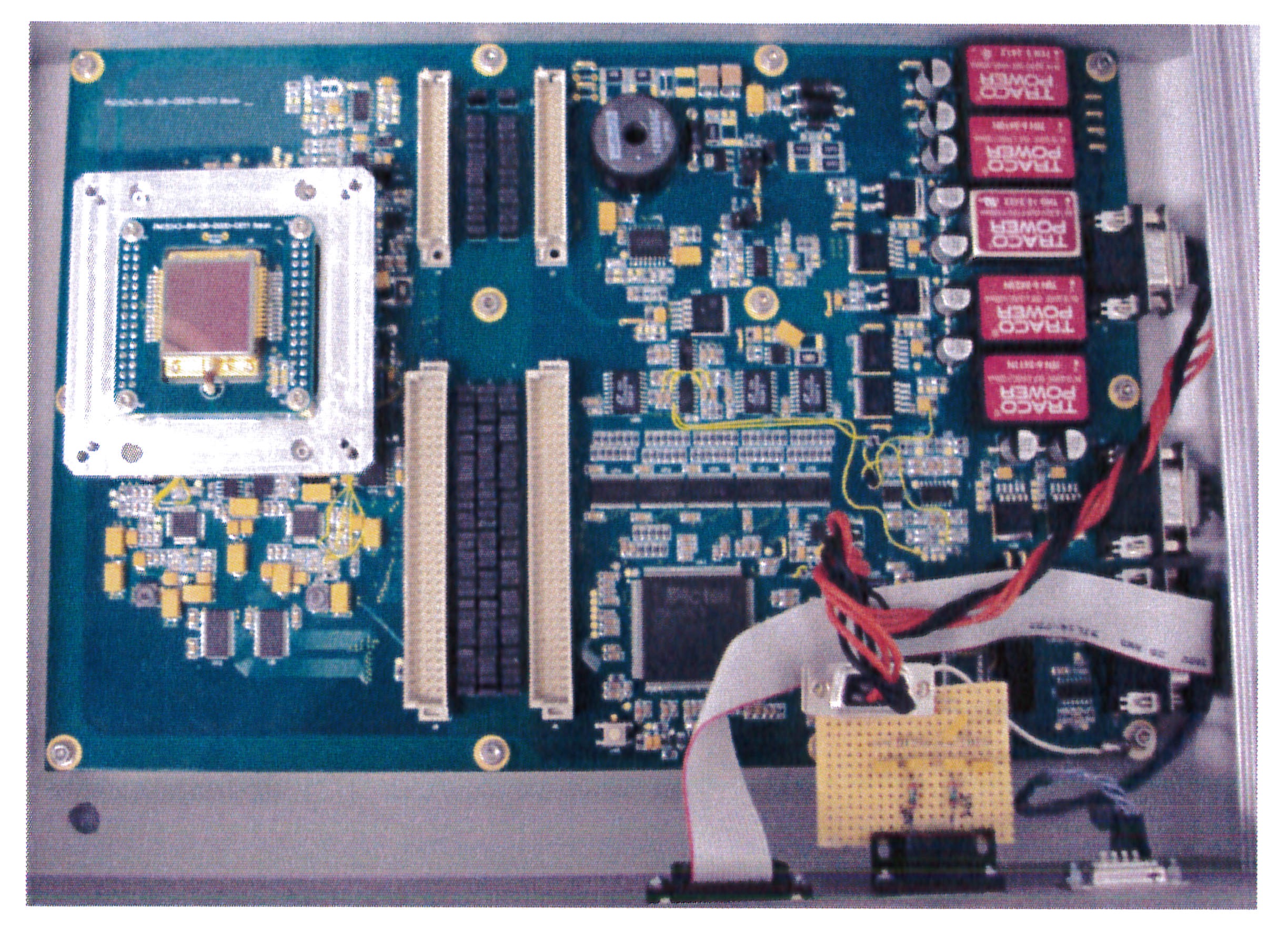}
\caption[IRCAM FEE prototype.]{IRCAM FEE prototype. \label{Fig:figure8}}
\end{center}
\end{figure}

Figure \ref{Fig:figure8} depicts the FEEP as delivered to IAC. The FEEP is hosted within a box that protects it mechanically. The box interfaces electrically with the main power supply, the trigger signal interface, and the ICU simulation by means of three different connectors, and allows assembling of the optics on top of the optical detector. 

For practical purposes, it has been decided to have a single board in which the FPA and FEE sections electrically isolated, and potentially joined by a connector with a similar length as the flight flex connector.
The FPA is located at the left half of the board, near the detector, while the FEE is hosted at the right half.

The detector is soldered on an independent assembly, being it possible to mount and dismount the detector PCB from the rest of the electronics. The detector assembly will consist of the following parts:
\begin{itemize}
\item	Optical detector.
\item	Heat conductor.
\item	Optical detector PCB.
\end{itemize}

The heat conductor has been designed to have the same volume as the flight one, and is placed in direct contact with the detector base. A metallic braided strap leads the heat from the heat conductor to a cold point external to the FEEP. The PCB has two connectors to provide the detector assembly with means to connect and disconnect it to/from the rest of the electronics.

In addition to the functional features of the flight design, the FEEP also has some other capabilities to ease the flight design testing and validation:
\begin{itemize}
\item	The FEE and FPA side can be electrically communicated by two different ways, namely, a harness simulating the flex connector in the flight design, or directly by using electrical connections that can be enabled or disabled by means of jumpers located on the PCB. This way the effect of the harness length on the image results can be quantified.
\item	Apart from the buck converter designed for feeding the TEC, the prototype also features a linear regulator with the same purpose. The use of a linear regulator increases power needs but decreases noise on power supplies, and therefore, a trade-off of the two technologies can be done.
\item	An external signal ($3.3$ V TTL or CMOS) allows triggering a single image acquisition.
\item	As already explained, the detector is mounted on an isolated assembly, and this might worsen NETD.
\item	A simulator of the temperature sensing within the optical detector is included to be able to test the temperature regulation circuit with a standard Peltier.
\item	A single voltage supply of $24$ V, $1$ A is needed to feed the FEEP, and some DC/DC converters and linear regulator generate internally the voltage required by the electronics to operate nominally.
\end{itemize}

\section{IRCAM FEE Prototype Verification}             

The main results, limitations and lessons learned derived from the test campaign performed at SENER facilities are presented in the next sections.

\subsection{Microbolometer Power Supplies Noise}             
The measurement has been done only in the frequency range from $10$ Hz to $51$ kHz with a Dynamic Signal Analyzer (DSA). Below $10$ Hz the measurement instrument, i.e. spectrum analyser, introduces noise one order of magnitude higher than the existing in the prototype board, and therefore, the lower frequency is limited to $10$ Hz. The maximum frequency the DSA can measure is $51$ kHz. A high pass filter with a cutting frequency of $1$ Hz has been used to eliminate the continuous signal at zero frequency. 

The power spectral density (PSD) of the noise of each power supply is summarized in Table \ref{Table:table4}.

\begin{table}[!h]
\caption[Power supplies noise PSD measurement.]{Power supplies noise PSD measurement.} \label{Table:power}
\begin{center}
\begin{tabular}{|p{40mm}|p{40mm}|}
\hline
\textbf{Power Supply}   & \textbf{Noise} \\
\hline
\hline
VSK@5V             &	 4.90925 $\mu$V$_{RMS}$ \\
\hline
VGSK               &	 4.98774 $\mu$V$_{RMS}$ \\
\hline
GFID@5V            &	 5.0095 $\mu$V$_{RMS}$ \\
\hline
VDDA             &	 8.5572 $\mu$V$_{RMS}$ \\
\hline
VDDL             &	 20.0827 $\mu$V$_{RMS}$ \\
\hline
GND             &	 4.91247 $\mu$V$_{RMS}$ \\
\hline
\end{tabular}
\end{center}
\end{table}

The noise in low frequency is the bigger contribution to the final noise and it decreases as frequency increases. To calculate the noise in the $10$ MHz bandwidth, the noise has been considered white noise and has been extrapolated, which is a worst case assumption. Using the spectral power distribution, the ratio between $10$ MHz and $51$ kHz is $14$, which leads to the PSD values shown in Table \ref{Table:table5}.

\begin{table}[!h]
\caption[Power supplies noise PSD extrapolation to $10$ MHz.]{Power supplies noise PSD extrapolation to $10$ MHz.} \label{Table:table5}
\begin{center}
\begin{tabular}{|p{30mm}|p{40mm}|p{40mm}|p{30mm}|}
\hline
\textbf{Power Supply}   & \textbf{Noise} & \textbf{Noise} & \textbf{Microbolometer} \\
   & \textbf{(10 Hz - 51 kHz)} & \textbf{(10 Hz - 10 MHz)} & \textbf{specification} \\
\hline
\hline
VSK@5V             &	 4.90925 $\mu$V$_{RMS}$ &	 68.743 $\mu$V$_{RMS}$ &	 100 $\mu$V \\
\hline
VGSK               &	 4.98774 $\mu$V$_{RMS}$ &	 69.842 $\mu$V$_{RMS}$ &	 100 $\mu$V \\
\hline
GFID@5V            &	 5.0095 $\mu$V$_{RMS}$ &	 70.147 $\mu$V$_{RMS}$ &	 100 $\mu$V \\
\hline
VDDA             &	 8.5572 $\mu$V$_{RMS}$ &	 119.824 $\mu$V$_{RMS}$ &	 100 $\mu$V \\
\hline
VDDL             &	 20.0827 $\mu$V$_{RMS}$ & 281.214 $\mu$V$_{RMS}$ &	 100 $\mu$V \\
\hline
GND             &	 4.91247 $\mu$V$_{RMS}$ &	 68.788 $\mu$V$_{RMS}$ &	 100 $\mu$V \\
\hline
\end{tabular}
\end{center}
\end{table}

All the values are compliant with the specifications with the exception of VDDA, which is slightly higher than specified. It could be due to the approximation done, which is a worst case consideration and in this case a measurement with higher bandwidth should be done.

\subsection{TEC long-term stability}             
In the preliminary design, a switching voltage regulator was chosen due to the better efficiency compared to the linear regulators. Nevertheless, the prototype has means to control the TEC in either a switching or a linear way, just by shorting or opening some jumpers.
One of the objectives of the FEEP fabrication is to decide which regulator configuration, switching or linear, is more suitable for JEM-EUSO needs. Our findings are the following:
\begin{itemize}
\item	For small thermal jumps, it is, difference between ambient and TEC temperature less than $10$ K, the linear regulator configuration is recommended, as the power needed to operate is small (less than $200$ mA at $3.3$ V, it is, $0.66$ W, for a $8^\circ$C thermal jump), and the switching regulator needs a minimum thermal jump to operate achieving the $10$ mK thermal stability requirement. Given the standard laboratory conditions in which the FEEP was tested, the linear regulator configuration was used to perform the acceptance test, although the switching configuration has also been tested and its performance is considered correct. 
\item	For big thermal jumps of more than $10^\circ$C, both configurations are possible to be used, although the linear configuration power consumption is bigger. In the tests performed at SENER in linear configuration, a thermal jump of $17$ K implied an increase of $3.3 \cdot 0.4=1.32$ W in power.
\end{itemize}

Due to hardware constraints in the design, the maximum allowed current to the TEC is around $1.2$ A, and therefore the maximum thermal jump allowed for the linear configuration is around $40-50$K. Nevertheless, for such big thermal jumps, the switching configuration is strongly recommended.

Depending on the final design of JEM-EUSO, either the switching or linear configuration, or even a combination of both of them, can be used.

At SENER, two tests have been carried-out to measure the long-term stability of the microbolometer temperature over $4.8$ hours, both in linear regulation configuration.

The first one with a temperature setting point of $30^{\circ}$C while the ambient is at $22^\circ$C showed a mean temperature of $30.0023^{\circ}$C and a standard deviation of $1.14$ mK, much lower than the $\pm 10$ mK required.

\begin{figure}[!tb] 
\begin{center}
\includegraphics[width=14cm]{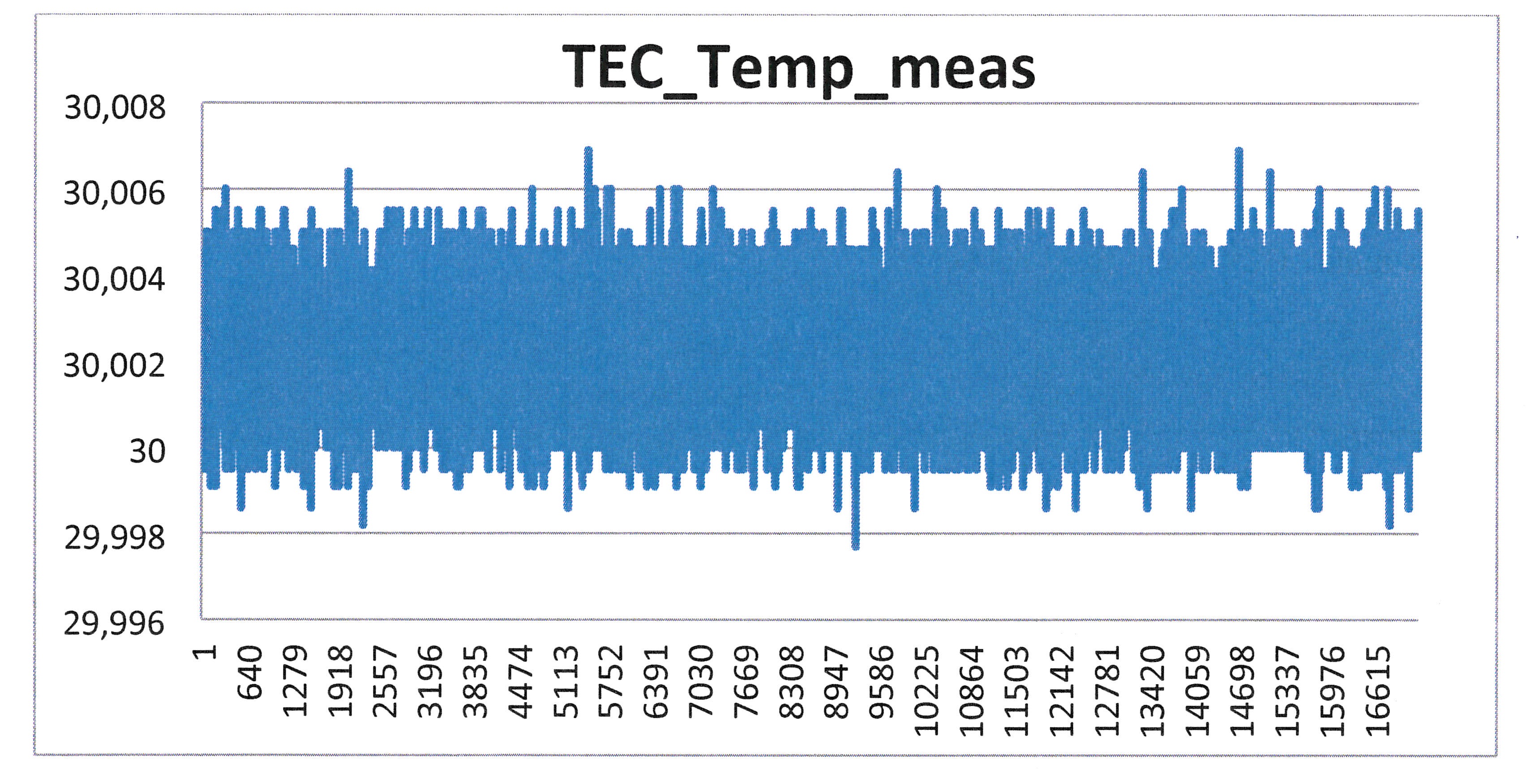}
\caption[Microbolometer temperature stability for 5th, $T_{amb}=22^\circ$C, $T_{sensing}=30^\circ$C.]{Microbolometer temperature stability for 5th, $T_{amb}=22^\circ$C, $T_{sensing}=30^\circ$C. \label{Fig:figure9}}
\end{center}
\end{figure}

The second test was executed similarly, changing the setting point to $39^\circ$C to have a $17^\circ$C difference between ambient and the setting point. In this case, the mean is $38.9966^\circ$C and the standard deviation $2.70$ mK, also within specifications.

\begin{figure}[!tb] 
\begin{center}
\includegraphics[width=14cm]{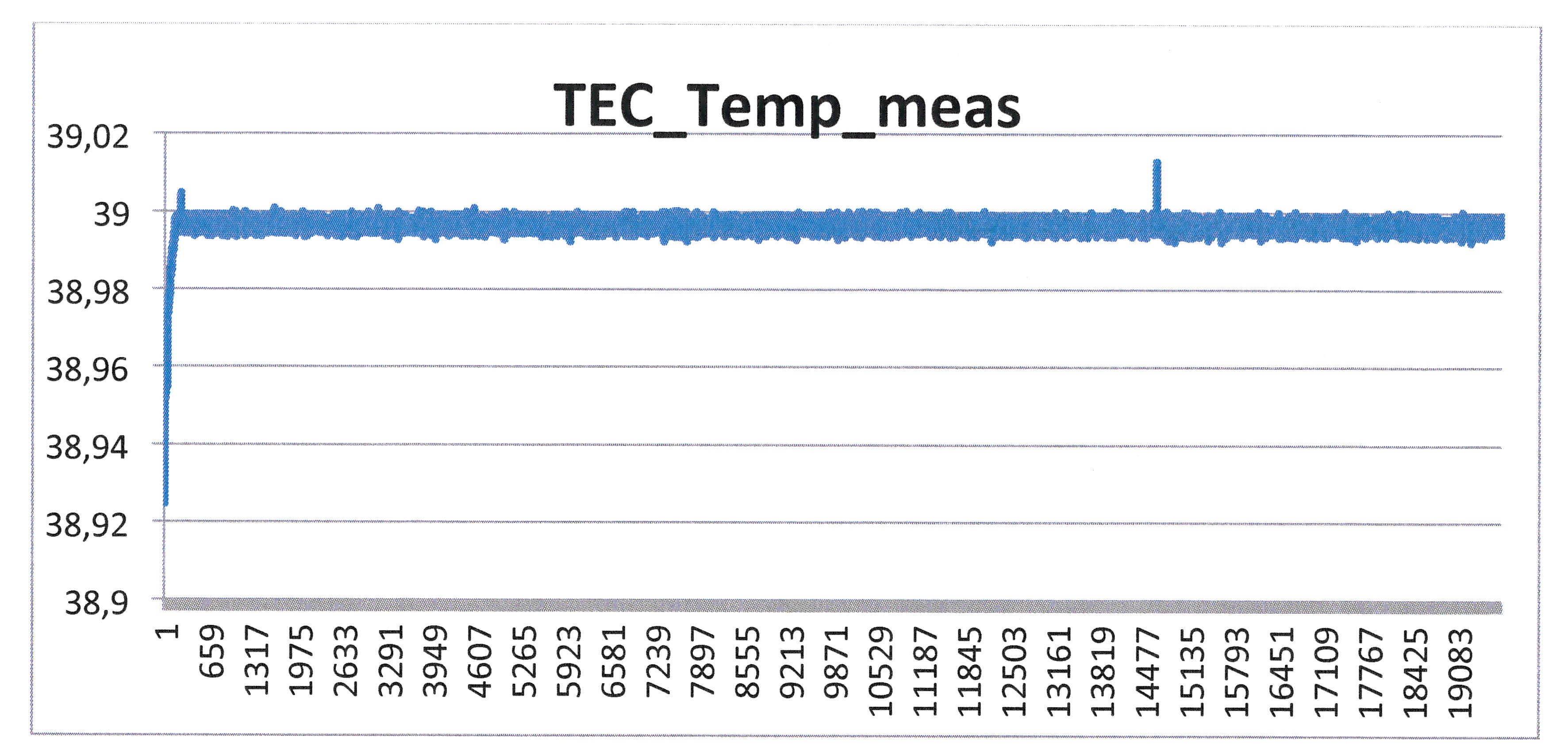}
\caption[Microbolometer temperature stability for $5$ h, $T_{amb} = 22^\circ$C, $T_{setting} = 39^\circ$C.]{Microbolometer temperature stability for $5$ h, $T_{amb} = 22^\circ$C, $T_{setting} = 39^\circ$C. \label{Fig:figure10}}
\end{center}
\end{figure}

\subsection{Dead pixel effect}             
Due to the big changes in microbolometer output voltage level during image transmission and the absolute maximum ratings for input voltage levels of the ADC, diodes have been added to protect the ADC. The use of the diodes and its slow recovery time imply every time a dead pixel is read, the effect in the output image is two consecutive pixels in the same channel seen as dead pixels. Alternatives to protect the ADC will be studied in further project phases.

\begin{figure}[!tb] 
\begin{center}
\includegraphics[width=14cm]{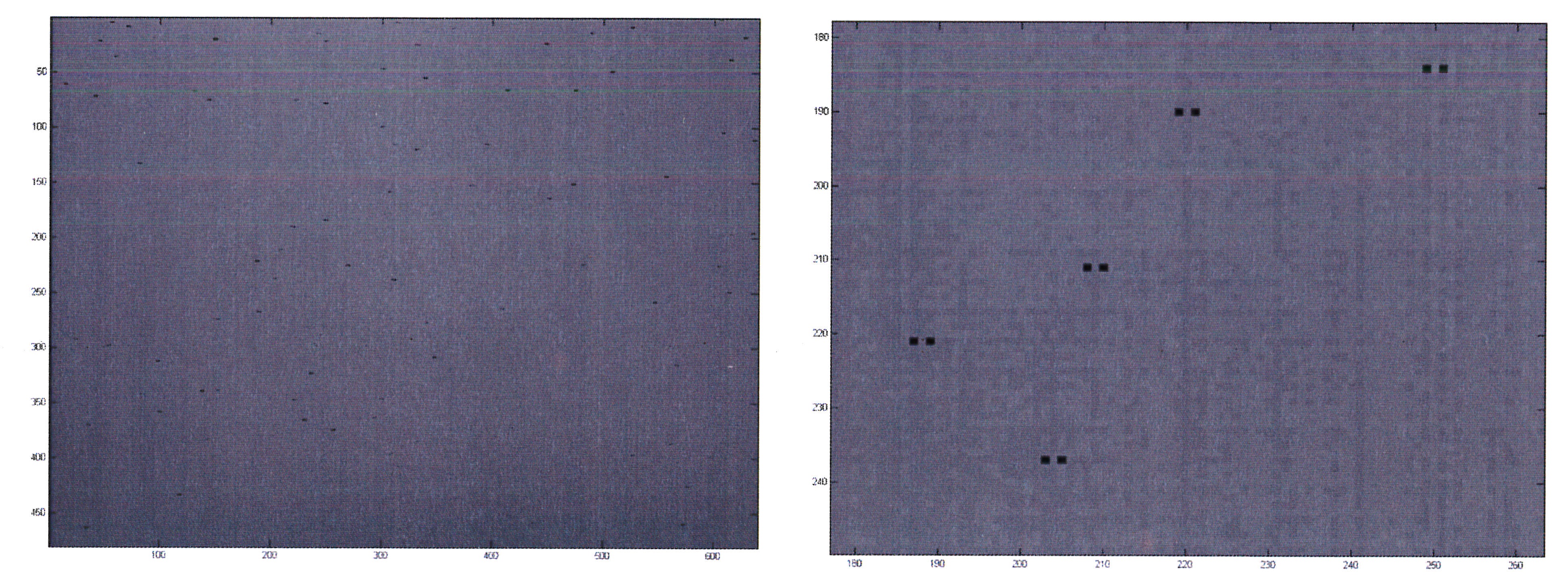}
\caption[Dead pixel map and detail of the double dead-pixel effect.]{Dead pixel map and detail of the double dead-pixel effect. \label{Fig:figure11}}
\end{center}
\end{figure}

\subsection{NETD measurement without microbolometer}             

The noise of the whole acquisition chain, both analog and digital contribution of the IRCAM FEE prototype without the microbolometer has been measured. An offset on both channels of $0$ V and a configurable input voltage generated by a voltage reference is used to acquire several complete images. Table \ref{Table:table6} shows the input voltage, the digital mean value and standard deviation obtained (in ADC counts and mK consideMicrobolomering a responsivity of $7.3$ mV/K) in each channel after the ADCs. 

\begin{table}[!h]
\caption[Noise measurement of the acquisition chains in the digital domain.]{Noise measurement of the acquisition chains in the digital domain.} \label{Table:table6}
\begin{center}
\begin{tabular}{|p{30mm}|p{30mm}|p{30mm}|p{30mm}|p{30mm}|}
\hline
\textbf{Input Voltage}   & \multicolumn{2}{c|}{Channel 1} & \multicolumn{2}{c|}{Channel 2} \\ \cline{2-5}
\textbf{[V]} & \textbf{Mean} & \textbf{NETD [ADC} & \textbf{Mean} & \textbf{NETD [ADC} \\
 & \textbf{[ADC counts]} & \textbf{counts, mK]} & \textbf{[ADC counts]} & \textbf{counts, mK]} \\
\hline
\hline
0.1            &	381.77 &	 0.95, 32 & 388.97	 & 0.82, 27	 \\
\hline
0.2            &	799.21 &	 0.94, 31 & 806.42	 & 0.8, 27	 \\
\hline
0.3            &	1206.10 &	 0.91, 30 & 1213.49	 & 0.8, 27	 \\
\hline
0.4            &	1618.38 &	 0.89, 30 & 1625.77	 & 0.74, 25	 \\
\hline
0.5            &	2031.47 &	 0.94, 30 & 2039.24	 & 0.74, 25	 \\
\hline
0.6            &	2439.02 &	 0.94, 31 & 2447.47	 & 0.74, 25	 \\
\hline
0.7            &	2853.08 &	 0.98, 33 & 2861.99	 & 0.75, 25	 \\
\hline
0.8            &	3265.37 &	 1.01, 34 & 3274.63	 & 0.76, 25	 \\
\hline
0.9            &	3677.69 &	 1.02, 34 & 3687.53	 & 0.77, 26	 \\
\hline
0.99           &	4048.44 &	 1.04, 35 & 4058.50	 & 0.77, 26	 \\
\hline
\hline
\end{tabular}
\end{center}
\end{table}

According to Table \ref{Table:table6}, channel $1$ is noisier than channel $2$. This could be due to the PCB routing, although no direct cause has been found. Note the maximum NETD is $35$ mK including also the reference voltage used to feed the FEEP, and therefore, this is a worst case NETD. Thus the worst case IRCAM FEE NETD, considering the microbolometer has $60$ mK NETD would be $69.46$ mK, which is below the specified $75$ mK. By using four images averaging, theoretically the NETD could be reduced by one half.

From Table \ref{Table:table6} it can also be seen that channel $2$ has $7$ counts offset with respect to channel $1$. This can be corrected by the FPGA as it has been characterized when a $0$ V input is inserted in both channel $1$ and $2$.

Another main result is that the NETD does not depend on the use of linear or switching regulation for the TEC control, so either regulator can be used in the final design.

These results can be complemented with the effects of the harness connecting FPA and FEE, which have not been evaluated so far. This is recommended to be done in the future to fully characterize the FEEP. 
 
\section{Conclusions} 
The IRCAM FEE preliminary design and prototype design, manufacturing and verification have been carried out. The overall results show the FEEP NETD is within NETD budget even without frame averaging.

The linear versus switching regulation options to control the microbolometer temperature have been studied. Accordingly, it is advisable to use linear regulation when the thermal jump between ambient and microbolometer temperature is below $10^\circ$C. Between $10^\circ$C and $20^\circ$C thermal jump, both linear and switching regulation options are possible, while above $20^\circ$C the switching option is strongly recommended.

Although the overall results are considered successful, the effect of the harness connecting FPA and FEE on the IRCAM FEE NETD should be characterized in future studies, and an alternative to the diodes protecting the channel acquisition ADCs causing the double dead pixel effect should be considered.

\section*{Acknowledgements} 
This work is funded by the Spanish Ministry of Economy and Competitiveness or MINisterio de Economia y COmpetitividad (MINECO) under projects AYA-ESP 2011-29489-C03-01, AYA-ESP 2011-29489-C03-02, AYA-ESP 2012-39115-C03-01, AYA-ESP 2012-39115-C03-03, CSD2009-00064 (Consolider MULTIDARK) and by Comunidad de Madrid (CAM) under project S2009/ESP-1496. 
We want to thank the JEM-EUSO collaboration, to which this work is entitled, and the Instituto de Astrofísica de Canarias (IAC) to give SENER the opportunity to join the JEM-EUSO project.



\begin{thebibliography}{9}
  \bibitem{Takahashi2009} Takahashi, Y. and the JEM-EUSO Collaboration, “The Jem-Euso Mission,” New J. Phys. 11(6), (2009).
  \bibitem{Ebisuzaki2014} 	Ebisuzaki, T., Medina-Tanco, G. and Santangelo, A.,“The JEM-EUSO Mission,” Adv. Sp. Res. 53(10), 1499–1505 (2014). 
  \bibitem{Adams2013}	J.H. Adams Jr. et al (JEM-EUSO Collaboration), “An evaluation of the exposure in nadir observation of the JEM-EUSO mission,” Astroparticle Physics, 44, 76–90 (2013) 
 \bibitem{Neronov2011} 	Neronov, A., Rodriguez-Frias, M. D., Toscano, S. and Wada, S., “Atmospheric Monitoring System of JEM-EUSO,” Proc. 32nd Int.Cosm. Ray Conf., 91–94 (2011). 
  \bibitem{Rodriguez-Frias2013} 	Rodriguez-Frias, M. D., Licandro, J., Sabau, M. D., Reyes, M., Belenguer,T., Gonzalez-Alvarado,M.C., Joven, E., Morales de los Rios, J. A., Saez-Palomino, M., Prieto, Saez-Cano, G., Carretero, J., Perez-Cano, S., del Peral, L., “Towards the preliminary Design Review of the Infrared Camera of the JEM-EUSO Collaboration” Proc. 33rd Int. Cosm. Ray Conf., 95–98, Rio de Janeiro (2013). 
  \bibitem{Morales2013} 	Morales de los Rios, J. A., del Peral, L., Saez-Cano, G., Prieto, H., Carretero, J. H., Sabau, M. D., Belenguer, T., Gonzalez Alvarado, C., Sanz Palomino, M. et al., “An End to End Simulation code for the IR-Camera of the JEM-EUSO Space Observatory,” Proc. 33nd Int. Cosm. Ray Conf., Rio de Janeiro (2013). 
  \bibitem{JEM-EUSO2014} 	The JEM-EUSO Collaboration, “The Infrared Camera onboard JEM-EUSO,” Experimental Astron., (in press, 2014). 
  \bibitem{Morales2014}	Morales de los Ríos, J. A., et al, “The infrared camera prototype characterization for the JEM-EUSO space mission,” Nuclear Instruments and Methods in Physics Research, 74–83 (2014)
  \bibitem{ULIS2014} ULIS, “Nano640E-UL 04 17 1-011,” 2014, in http://www.ulis-ir.com/index.php?infrared-detector=25--\%m-640x480, (23 April 2014).
\end{thebibliography}
\end{document}